# Opto-twistronic Hall effect in a three-dimensional spiral lattice

Zhurun Ji[1,2], Yuzhou Zhao[3], Yicong Chen[2], Ziyan Zhu[4], Yuhui Wang[1], Wenjing Liu[1,5], Gaurav Modi[1], Eugene J. Mele[2], Song Jin[3] and Ritesh Agarwal[1]

1 Department of Materials Science and Engineering, University of Pennsylvania, Philadelphia, PA 19104, USA

2 Department of Physics and Astronomy, University of Pennsylvania, Philadelphia, PA 19104, USA.

3 Department of Chemistry, University of Wisconsin-Madison, Madison, WI 53706, USA

4 Stanford Institute for Materials and Energy Sciences, SLAC National Accelerator Laboratory, Menlo Park, CA 94025, USA

5 State Key Laboratory for Mesoscopic Physics and Frontiers Science Center for Nano-optoelectronics, School of Physics, Peking University, Beijing, 100871, China

**Studies of moiré systems have elucidated the exquisite effect of quantum geometry on the electronic bands and their properties, leading to the discovery of new correlated phases. However, most experimental studies have been confined to a few layers in the 2D limit. The extension of twistronics to its 3D limit, where the twist is extended into the third dimension between adjacent layers, remains underexplored due to the challenges in precisely stacking layers. Here, we focus on 3D twistronics on a platform of self-assembled spiral superlattice of multilayered $WS_2$. Our findings reveal an opto-twistronic Hall effect driven by an**

**interplay of structural chirality and photon momentum, modulated by the moiré potential of the spiral superlattice. This coupling is an experimental manifestation of the noncommutative geometry that arises when simple translational symmetry perpendicular to the stack is replaced by a non-symmorphic screw operation. We also discover signatures of enhanced light-matter interactions and a strongly coupled photon momentum-lattice interaction owing to the properties of the 3D twistronic system that cannot be explained by existing theoretical models. Crucially, our findings uncover the role of higher-order quantum geometrical tensors in light-matter interactions, opening new avenues for designing quantum materials-based optoelectronic lattices with large nonlinearities, paving the way for developing advanced nanophotonic devices.**

Chiral states of matter are known to provide interesting functionalities. Some of the most interesting molecular structures discovered include the double-helical DNA and triple-helical collagen and extensive research has elucidated intriguing structure-property relationships in materials synthesized by their hierarchically ordered assemblies[1, 2, 3]. Analogously, one-dimensional helical nanotubes[4, 5] and two-dimensional (2D) materials[6, 7] assembled via control of their twist angles (moiré systems) have been widely explored, and the effect of geometry on electronic properties of quantum matter as well as the emergence of new phases have attracted attention[8]. Linear and nonlinear optical response studies have also uncovered new properties associated with the modified lattice geometry in 2D moiré systems such as interlayer exciton emission[9] and mediated interaction[10], quantum anomalous Hall effect[11, 12], intelligent infrared sensing[13], and moiré exciton-polaritons[14]. However, most of these studies have been limited to twisted few layer homo- and hetero-bilayer transition metal dichalcogenide (TMD) systems, hence mostly confined to two-dimensions. This is primarily due to the challenge of manually stacking individual layers practically constraining most studies to only a few layers[15, 16, 17]. However, it is natural to ask if the idea of twistronics can be continued into the third dimension and if new phases and properties can emerge in these 3D chirally stacked systems. There have been a few theoretical predictions on the possible changes to the system when the layer numbers grow, such as moiré of moiré modulation[18], topological phase transition[19], optical rotation[20], and the emergence of new resonant states[21, 22]. However, these properties have not been experimentally explored in systems where the twist forms a pattern or (quasi)-periodicity in the third dimension, and connects the 2D moiré systems with the multilayered, 3D limit, i.e., solid state analogs of helical DNA.

In few-layer twisted 2D moiré materials, electron (exciton)-photon interactions have been shown to share many characteristics with that of monolayers. Therefore, in the 3D limit of moiré systems, an intriguing question arises when the moiré lattice periodicity and sample thickness become comparable to the optical wavelength scale: how is the light-matter interaction paradigm affected? This question is important since the electrodynamic response of engineered quantum materials reveals unique feature of their electronic states and that are not ordinarily studied in systems where lattice and optical length scales become comparable[23, 24]. Furthermore, for conventional 3D crystals that are characterized by three mutually commuting primitive translation vectors, momentum space cleanly separates into invariant sectors characterized by a conserved crystal momentum. The situation is more complex for systems where the primitive translations do not commute. A famous example occurs for the motion of charged quantum particles in a uniform magnetic field but noncommutative translations can be induced by appropriately constraining the dynamics by a projection onto a restricted subset of bands. Since supertwisted lattice has the noncommutative geometry, its mesoscopic responses could be different from conventional lattices.

Also, with large length scales of a moiré system, it is expected to exhibit strong light-matter coupling, enhancing optical nonlinearities and reveal novel responses associated with their underlying quantum geometry[25], opening up many possibilities of designing innovative quantum photonic devices.

To study the properties of 3D chirally stacked systems, we investigate supertwisted spirals of $WS_2$ formed by a screw-dislocation-driven growth mechanism in a non-Euclidean geometry[26] via a direct chemical vapor deposition growth (Fig. 1a) (see Methods). These self-assembled 3D supertwisted samples consist of tens to hundreds of $WS_2$ layers with a nearly uniform twist angle between layers and display a nonsymmorphic screw symmetry, akin to a DNA strand albeit in the solid-state form (Fig. 1b). Further, scanning transmission electron microscopy and second harmonic generation results revealed the presence of moiré of moiré superlattices in supertwisted $WS_2$ samples due to the interference of moiré patterns[26, 27]. When layers of $WS_2$ are chirally stacked on top of each other with a small twist angle, we expect a plethora of high-order moiré patterns of different length scales, which are orders of magnitude larger than the length scale found in typical bilayer moiré systems[18] (Fig. 1c). We consider specifically the type of hexagonal supertwisted $WS_2$ samples[26, 28] where the base is 2H stacked $WS_2$ (Fig. 1a). The unique geometry of the supertwisted lattice with many degrees of freedom, including chirality, layer number, twist angle and associated moiré potential distribution (analogous to DNA's base, length, pitch, and supercoiling), poses an extreme challenge to theoretical analysis. Since the presence of a large number of layers in the system congests the linear optical spectra (Fig. 1c; discussed later) thereby limiting the information content, we performed photon momentum-dependent nonlinear optical Hall measurements as a spectroscopic tool to gain a microscopic understanding of fundamental optical excitations in these 3D chiral systems and explore their potential for studying unconventional quantum phenomena. Our observations unveil a unique opto-twistronic Hall response driven by the noncommutative geometry of the supertwisted chiral system highlighting the interplay between structural, electronic, and quantum geometrical factors, demonstrating their versatility in chiral photovoltage generation.

First, we investigated the linear optical properties of the supertwisted $WS_2$ spirals at room temperature. Notably, we observed a distinct intralayer A exciton reflectance dip at ~636 nm (Fig. 1c), with no discernible differences in the reflectance spectra due to the structural chirality of the system when interacting with left and right circularly polarized light, above the detection limit [29]. (Fig. 1c, details in SI note 1.2, Fig. S2). The schematic in Fig. 1d illustrates the exciton formation process; in a multilayer supertwisted $WS_2$ stack, exciton eigenfunctions are composed of an in-plane component and a standing wavefunction in the $z$-direction[30]. It distinguishes itself from 2H- or 3R- stacked $WS_2$[17, 28], because the lattice twist gives rise to a series of shifted eigenmodes ($t$ is the interaction strength). When the stack becomes thick, the energy separation between neighboring modes becomes smaller than spectral broadening and the resonances overlap to form a single broad dip, as observed. Therefore, the supertwisted samples call for a different optical technique to explore their properties originating from spatial and geometrical degrees of freedom.

To further investigate this 3D chiral system, we employed optical Hall measurements, which is a nonlinear probe where the magnetic field in the conventional Hall effect is replaced by a polarized laser beam and is extremely sensitive to the symmetries of the system (see Methods) [31, 32]. For these measurements, the supertwisted $WS_2$ flakes were etched to into a Hall bar geometry with uniform thickness (Fig. 2a, device D1; see Methods). Since the samples show strong excitonic features, upon optically exciting the sample near the exciton resonance wavelength, an in-plane voltage $V_0$ was applied to dissociate the charge-neutral electron and hole pairs, and the generated photovoltage $V_H$ was measured along the perpendicular in-plane direction as a function of optical polarization (see Methods). In our experimental geometry, the laser was incident on the sample at a zenith and azimuthal angles of $\theta$ and $\alpha$, respectively, passing through a quarter wave plate (QWP), with a rotation angle $\phi$ with respect to its fast axis (Fig. 2b, also see SI Note 1.1). In device D1, when applying $V_0 = 1\,V$, the photovoltage plot was recorded as a function of QWP angle (laser incident normally to the sample surface, i.e., $\theta = \alpha = 0$ ). The red line in the middle panel (Fig. 2c shows the part sensitive to the circular polarization of light, or circular Hall photovoltage, $V_{HC}$, and blue line in the lower panel shows the linear polarization-dependent Hall photovoltage, $V_{HL}$. The results show that Hall photovoltage depends on both the linear and circular polarizations. To examine the possible origins of $V_{HC}$, the applied transverse voltage $V_0$ was swept while $V_{HC}$ were measured at different excitation wavelengths (see Methods). The observed linearity of both $V_{HC}$ and longitudinal current $I_0$ response as a function of the applied $V_0$ (Fig. 2d) shows that the fabricated bismuth/gold electrodes[33] formed ohmic contacts with the $WS_2$ flakes, which minimizes the Schottky junction contributions to our measurements[34]. Therefore, these results suggest an intrinsic Hall response in the supertwisted system that is sensitive to light chirality. In contrast, the optical Hall response vanishes in 2H-stacked multilayer $WS_2$ systems[35] due to the lack of mechanisms that can separate the trajectories of electrons and holes when excited with light with different circular polarization. This chiral light-dependent photovoltage response unique to supertwisted systems enables us to go beyond the limitation of linear optical conductivity and study the interlinkage between symmetry and geometrical properties in the system, as discussed below.

To further explore the influence of the handedness of the structural stacking of supertwisted $WS_2$ on its nonlinear optical response, we measured the circular polarization Hall photovoltage, $V_{HC}$, on a series of samples in the Hall bar geometry (Figs. 3a, b). Since the optical Hall response ($V_{HC}$) vanishes when the interlayer twist angle is 60° (i.e., 2H stacking), we selected two supertwisted WS$_2$ samples to study the onset of optical Hall voltage with supertwisting angle: device D2 with counterclockwise (~ 8° twist between adjacent layers) and device D3 with clockwise (~ −8° between layers) twisting from the bottom to the top layer (see SI note 1.5, Fig. S10 for optical and AFM images). With a similar device geometry and measurement configuration, the optical Hall resistance $R_{HC}$ (obtained from the measured photovoltage $V_{HC}$) shows a sign reversal (Fig. 3a, b) with opposite helicities of the supertwisted sample, showing that the handedness of the structural supertwisting governs the optical Hall response, and hence we name it the *opto-twistronic Hall effect.*

To describe the opto-twistronic Hall phenomena in the 3D chiral system, we model the light field as $E_i(r,t) = E_i^{i(q \cdot r - \omega t)} + c.c.$ where $q$ and $\omega$ span the momentum and frequency spaces. The circular polarization dependent Hall current $J(q, \omega)$ is determined by a nonlinear conductivity tensor $\sigma^{(2)}$, and $(E \times E^*)$, the pseudo-vector depicting light helicity. $\sigma^{(2)}$ is a material dependent, complex function in the $(\omega, q)$ space. For the case of a monolayer WS$_2$ under bias, $\sigma^{(2)}$ originates from the opposite Lorentz-like forces experienced by the electrons from different valleys, i.e., the valley Hall effect[35]. In the case of 2H- stacked multilayers, $\sigma^{(2)}$ in each layer is cancelled out by the neighboring layer due to valley degeneracy.[36] In contrast, carrier dynamics in the supertwisted material can be analyzed by considering the free propagation of a particle in a helical frame of reference. A free particle Hamiltonian quadratic in momentum, $H = p \cdot p / 2m$ in the co-rotating frame of the crystal is augmented by a geometric connection as seen in the lab frame, leading to a quasi-momentum operator, $\hbar \kappa_z = p_z + \hbar \beta \widehat{m}_z$, where $\beta =$ $d\phi/dz$ is the pitch of the helix and $\hat{m}_z = xp_y - yp_x$ is the $z$-component of the orbital angular momentum. $\kappa_z$ is the generator for the nonsymmorphic screw symmetry operation of the stack and its eigenvalues play the role of conserved Bloch momenta for interlayer translations coupled to rotations. Completing the squares in $H$ gives the transformed Hamiltonian,

$$H = \frac{1}{2m}\left[(p_x + \hbar\kappa_z\beta y)^2 + (p_y - \hbar\kappa_z\beta x)^2 + (\hbar\kappa_z)^2(1 - \beta^2(x^2 + y^2)) + (\hbar\beta\hat{m}_z)^2\right] \quad (1)$$

with transformed momenta $\Pi_x = p_x + \hbar\kappa_z\beta y$ and $\Pi_y = p_y - \hbar\kappa_z\beta x$ that obey the commutation relation $[\Pi_x, \Pi_y] = 2i\hbar^2\kappa_z\beta$. Thus, in each sector with fixed $\kappa_z$, the in-plane differential translations do not commute, completely analogous to the situation for a particle with charge $e$ moving in a uniform magnetic field with strength $B^* = 2\hbar\kappa_z\beta/e$. However, unlike the situation for a physical magnetic field, here the effective field $B^*$ is odd in the quasi-momentum $\kappa_z$, reflecting the fact that the twisted structure preserves time reversal symmetry. Any nonequilibrium state distribution that breaks time reversal symmetry, e.g. under excitation with circularly polarized light, will undergo a net transverse deflection in response to an applied in-plane bias and can be identified by the induced Hall response. Importantly, field responsible for this deflection $B^*$ is also odd in $\beta$, so that the sign of the Hall field is reversed in enantiomeric structures, as experimentally observed. The effective field $B^*$ is now a purely geometrical quantity, originating from its noncommutative property and specified by the degree of twist and the conserved value of $\kappa_z$. We estimate $B^*$ to be ~ 4 T for the supertwisted $WS_2$ structures with ~8° twist angle.

To explore how the properties of the twisted chiral stack evolves with thickness, $V_{HC}$ was systematically measured on a thick supertwisted $WS_2$ flake at different sample thicknesses regions (i.e., stacking layer number; locations denoted by green circles on the device D4 image shown in Fig. 3c inset) at 77 K. In the thinnest region, i.e., P1 location in Fig. 3c, the $V_{HC}$ spectrum (i.e., laser excitation energy dependence of $V_{HC}$) shows a prominent peak at the A exciton wavelength (~620 nm) and another smaller, redshifted peak associated with the charged exciton absorption (A-)[37]. As the sample thickness increases, the $V_{HC}$ peak splits and the splitting increases monotonically with sample thickness. These observations show that the $V_{HC}$ spectrum evolves with features that are not simply a superposition of responses from different layers. The splitting of the $V_{HC}$ peak with increased thickness is consistent with the stronger light matter coupling observed in the reflectance spectra of strongly coupled optical cavities[38] but in our measurements, it is seen in the circular polarization dependent $V_{HC}$ data and without any external cavities. Angle-resolved linear reflectance measurements were also conducted to confirm the exciton-photon mode coupling[38, 39, 40] in the samples (details in SI, Note 1.3). These results suggest that the exciton-photon coupling in the supertwisted spirals strongly affects the nonlinear optoelectronic response from the hybridized states along with a strong chiral response from structural handedness. It also demonstrates that upon stacking from the 2D to 3D limit along with the nonsymmorphic rotation axis, the system develops another controllable degree of freedom, i.e., the number of layers, which has a significant impact on carrier excitation and transport and hence the spectra of the nonlinear optical Hall effect.

Both the layer thickness and handedness dependence of the opto-twistronic Hall effect suggest its relationship with the supertwisted system's nonsymmorphic symmetry, which contributes to modifying both its in-plane and out-of-plane electronic structure. To further explore the role of the pitch of the helix (i.e., twist angle between the layers) on the nonlinear optical response, we altered the direction and magnitude of photon momentum (projected onto the supertwisting axis ($\hat{z}$)) and mapped the changes in $V_{HC}$. Measurements were carried out on circular shaped supertwisted $WS_2$ devices, i.e. a symmetric Hall bar geometry (device D5, Fig. 4a; sample details in SI note 1.4), where $V_{HC}$ was measured as a function of the $\theta_y$ (oblique incidence angle in the y-direction; see schematic in Fig. 2c) at the A exciton wavelength (620 nm). Laser incidence angle-dependent photogalvanic effect measured in quantum wells, monolayer TMDs and other systems were typically observed to be sinusoidal, which reveals the dominant matrix elements of photogalvanic conductivity and related sample symmetries[37, 41]. As shown in Fig. 4b inset, the supertwisted WS$_2$ sample D6 has ~ -15° to -20° twist angles between neighboring layers, and the incidence angle-dependence $V_{HC}$ was observed to be sinusoidal, peaking at normal incidence, similar to the case observed in monolayer TMDs[37]. On the other hand, sample D5 with similar thickness and geometry but smaller twist angles between the neighboring layers (< 5°; Fig. 4c inset) shows a strongly non-sinusoidal angle dependence. To examine whether this unusual incidence angle dependence is a result of sample inhomogeneities, strain or other external factors, the spatial distribution of $V_{HC}$ was mapped out (Fig. S9). Since the results did not vary significantly across the circular flake, such extrinsic factors cannot explain the observed phenomena. A Fourier decomposition of the $V_{HC}$ curve (green dotted fitting curve in Fig. 4c) suggests that a significant contribution from higher order terms (i.e., $\sin^2 \theta$, $\sin^3 \theta$) is present, which implies contributions from the higher order photon momentum terms.

The strong non-sinusoidal feature observed in the light incidence angle-dependent $V_{HC}$ measurements imply a complex dependence of the opto-twistronic Hall voltage on photon momentum ($q$). This is different from the conventional second order optical conductivity, i.e., $\sigma^{(2)}(q=0,\omega)$ in the long wave approximation, or optical processes like photon drag[42] and spatially dispersive photogalvanic effects[43, 44], which are described by the first order in photon momentum $\sigma^{(2)}(q^1,\omega)$ conductivity tensor. Therefore, to describe the strong non-sinusoidal, higher order components of $V_{HC}$ in the supertwisted sample with smaller twist angles, a more general momentum space dispersion of the response needs to be considered, i.e., a generalized conductivity tensor $\boldsymbol{\sigma}^{(2)} = \boldsymbol{\sigma}^{(2)}(\boldsymbol{q}, \omega)$. Using the density matrix formalism[45], we derive the response function of the second order photoconductivity $\boldsymbol{\sigma}^{(2)}(0; \omega, -\omega)$ (see SI note 2.3 for a detailed derivation). In the nonlinear conductivity expression, $\boldsymbol{\sigma}^{(2)}$ is related to, $\tilde{v}_{kk'nm} = \frac{1}{2}\langle u_{n,k}|\{v, e^{iq\cdot r}\}|u_{m,k'}\rangle$, the generalized velocity operator which is nondiagonal in $\boldsymbol{k}$, between two Bloch states at $\boldsymbol{k}$ and $\boldsymbol{k}'$. Since $\tilde{v}_{kk+qnm}$ is evaluated between two closely spaced momenta $\boldsymbol{k}$ and $\boldsymbol{k+q}$, it provides a direct measure of the momentum dependence of the electronic Bloch wave functions, encoded in the band geometry (as the schematics in Fig. 4d-f illustrate). The cross product of $\tilde{v}_{kk'nm}$ in the response function is related to the band resolved Berry curvature, $\Omega^i_{nm}(\boldsymbol{k})$, as $\Omega^i_{nm}(\boldsymbol{k}) = -\frac{i}{(\epsilon_{km}-\epsilon_{kn})^2}(\tilde{v}^j_{kk'nm}\tilde{v}^k_{kk'mn} - \tilde{v}^k_{kk'nm}\tilde{v}^j_{kk'mn})$ when $\boldsymbol{k}' = \boldsymbol{k}$, and can be expanded as its multipoles when considering any general $\boldsymbol{k}$ (or $\boldsymbol{k}'$).

To understand the unusual photon momentum dependence of Hall photovoltage in the supertwisted system, we performed numerical simulations of $\sigma^{(2)}_{xxxy}$ ($\boldsymbol{q} = q_x\hat{x}, \boldsymbol{J} = j_x\hat{x}, \boldsymbol{E} \times \boldsymbol{E} = (E_xE_y - E_yE_x)\hat{z}$) on a few different continuum models of TMD stacks (details in SI note 2.4). Here excitonic effects and dc electric field are not incorporated for the sake of computational accessibility, but without loss of generality (also see SI note 2.1). The calculation uses a three-band tight binding model of monolayer TMD[46] (Fig. 4d) to show that $\sigma^{(2)}_{xxxy}$ does not have any dependence on $q_x$, as governed by $D_{3h}$ symmetry selection rules. This validates the usually adopted long wavelength approximation in conventional lattices. In comparison, $\sigma^{(2)}_{xxxy}$ calculated for a K-valley twisted homobilayer TMD model system[47] with a 1° twist angle and a moiré length ~10 nm (the scaling factor is $\frac{1}{\beta}$, $\beta$ being the twist angle, Fig. 4e) reveals that $\sigma^{(2)}_{xxxy}$ - $\boldsymbol{q}$ relationship is linear, where higher order of $q$ terms have non-negligible contributions to $\sigma^{(2)}_{xxxy}$. In the third model system, $\sigma^{(2)}_{xxxy}$ is calculated for a supertwisted TMD[48] model with 2° twist angles (in Fig. 4f, the red dotted line is a polynomial fitting). Like the effective model in Eqn. (1), the microscopic continuum model considers the noncommutative geometry of the supertwisted system. The fundamental unit of the model is a 4-band trilayer TMD model [47] that is coupled to the neighboring layers in the top/bottom block, which allows us to perform a Fourier expansion along the $z$ direction and define a smaller Brillouin zone for the trilayer system. We average $\sigma^{(2)}_{xxxy}$ over the wavevector along the z-direction, $k_z$, which shows a very prominent collective contribution from higher order $q$ terms (Fig. 4f). Comparing with the twisted bilayer model, this calculation shows an opposite sign of $\sigma^{(2)}_{xxxy}$ due to the dominant contribution from supertwisting, providing further evidence of the relationship between the noncommutative geometry and the observed opto-twistronic Hall effect. Although the model does not include all the ingredients of the system, the connection between the large moiré length scale in small angle supertwisted sample and the nonlinearity in $q$ terms in the response is also qualitatively consistent with the experimental results. In supertwisted TMDs, the variation along the $z$ direction can have different length scales, including large length scales that are comparable to the optical wavelength. Averaging these different scales can alter the optical Hall effect and amplify contributions from higher order of $q$.

Phenomenologically, the smaller twist angle will enhance the interband response from higher order changes of the Bloch wavefunctions (Fig. 4f), reflected in a strongly amplified $k$ to $k+q$ momenta correlation, and a distinct $k$-length scale for the Berry connection in the moiré of moiré bands[49]. Each higher (in photon momentum, $q$) order of conductivity component has a corresponding gauge invariant geometric quantity (SI note 2) that relates to multipolar excitation processes (similar to the case of "shrunken" light discussed in the context of plasmonics[23]), and each order can have comparable contributions. While the lowest order band geometrical tensor[25] has distinctively featured a wide range of quantum phenomena, including quantized anomalous hall effect[50] driven by magnetization/Berry curvature integral (Fig. 4g), quantized circular photogalvanic effect in topological semimetals[51], and nonlinear Hall effect[52, 53, 54] (Fig. 4h) caused by Berry curvature dipoles, our results demonstrate the opto-twistronic Hall effect (Fig. 4i) as a quantum phenomenon involving higher multipoles of band resolved Berry curvature. Since moiré systems are closely related to strong correlation physics and nontrivial band topology, our nonlinear optoelectronic probes provide a promising approach to obtain new insights into complex systems to unravel the interplay between quantum geometry, topology and strong electronic correlations.

In conclusion, our study unveils profound modifications in the interactions between light and matter within a chiral, three-dimensional moiré system comprised of supertwisted $WS_2$ spiral structure.s The noncommutative geometry of the system gives rise to the signature opto-twistronic Hall effect with the response determined by the handedness of the twisted layers and helicity of excitation light. As the potential modulation length scale in these 3D moiré systems approaches the scale of optical wavelengths, we observed a strong alteration of the nonlinear optical response from photon momentum, offering new insights and prompting further theoretical efforts into chiral optical properties and quantum geometry measurables pertaining to higher-order band-resolved Berry curvature. Moreover, the unique characteristics of supertwisted spiral systems serve as a vital link between 2D and 3D twistronics, overcoming the substantial length-scale differences between electrons and photons to induce enhanced light-matter interactions and significant optical nonlinearities. This versatile 3D supertwisted material platform, serving as both a quantum and photonic crystal, in conjunction with its opto-twistronic effect, not only expands our understanding of light-matter interactions but also opens exciting avenues for the development of novel nonlinear quantum photonic devices, pushing the boundaries of light-induced quantum phenomena.

**Acknowledgments**

This work was supported by the US Air Force Office of Scientific Research (award# FA9550-20-1-0345), National Science Foundation (NSF-QII-TAQS-#1936276 and NSF- 2323468) and Office of Naval Research (via Grant No. N00014-22-1-2378). This work was partially supported by the King Abdullah University of Science & Technology (OSR-2020-CRG9-4374.3) and NSF through the University of Pennsylvania Materials Research Science and Engineering Center (MRSEC) (DMR-1720530) seed grant. Device fabrication and characterization work was carried out in part at the Singh Center for Nanotechnology, which is supported by the NSF National Nanotechnology Coordinated Infrastructure Program under grant NNCI-1542153. Work by EJM is supported by the Department of Energy under Grant DE-FG02-84ER45118. Z.J. and Z. Zhu acknowledge support from the Stanford Science fellowship.

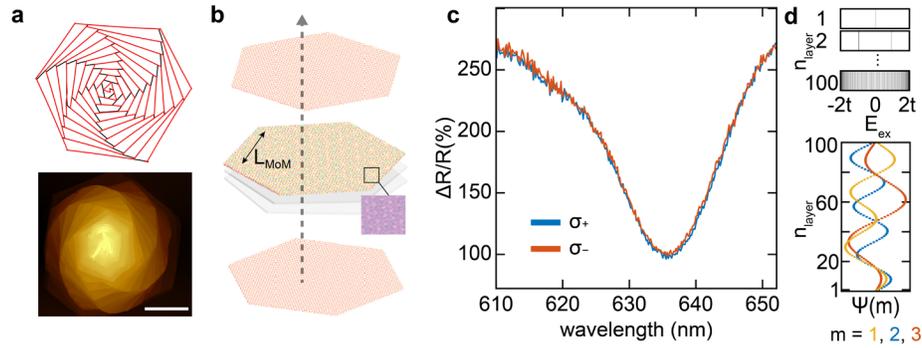

**Fig. 1 Illustration and linear optical characterization of the 3D chiral systems from supertwisted spirals of WS$_2$. a.** Top: Schematic of the supertwisted sample based on hexagonal WS$_2$ spirals (2H stacking arrangement). Bottom: AFM image of a characteristic supertwisted WS$_2$ sample with continuously twisting layers. Scale bar: 4 μm. **b.** Illustration of the moire of moire pattern in the supertwisted system. The grey arrow shows the screw rotation axis. L$_{MOM}$ denotes for the length of a moire of moire unit cell, which can be orders of magnitudes larger than the crystal unit cell and can become comparable to optical wavelengths (~100 nm). The zoomed in illustration shows the moire pattern formed in the bilayer region with a smaller lengthscale than that in the multilayers. **c.** Reflectance spectra measured with left and right circularly polarized light. **d.** Illustrations of the exciton eigenenergy distributions (top, where the distributions for 1, 2, and 100 layers systems are shown), and layer-hybridized exciton wavefunctions (m=1-3 in a 100-layer system) in the supertwisted system (bottom).

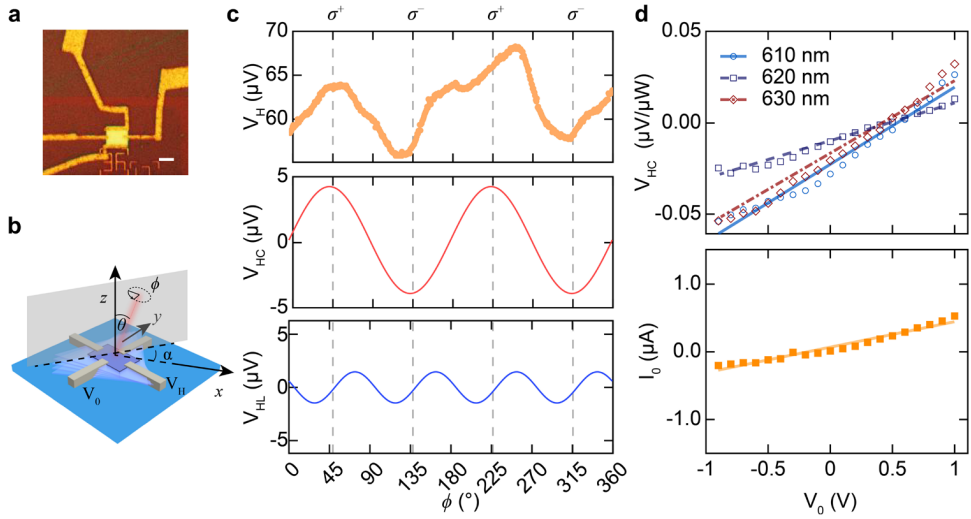

**Fig. 2 Opto-twistronic Hall effect measurements on the supertwisted WS$_2$ sample. a.** Optical image of the Hall bar device (D1) based on the supertwisted WS$_2$ sample, fabricated with Bi/Cu electrodes. Scale bar: 5 $\mu$m. **b.** Schematic of the optoelectronic measurement with incidence zenith angle $\theta$, azimuthal angle $\alpha$, quarter wave plate rotation angle $\phi$ with respect to its fast axis, applied transverse voltage $V_0$ and measured Hall voltage $V_H$. **c.** Top: Hall voltage $V_H$ as a function of $\phi$ at $V_0 = 1\,V$ is shown as the set of yellow dots. Middle: The red line is a fitting for the circular polarization dependent part of $V_H$, or $V_{HC}$; Bottom: The blue line is a fitting for the linear polarization dependent part of $V_H$, or $V_{HL}$. **d.** Top: Measurement of transverse $V_{HC}$ as a function of applied voltage $V_0$ at three different excitation wavelengths, showing an ohmic behavior. Bottom: Longitudinal photocurrent $I_0$ as a function of $V_0$ also showing ohmic response.

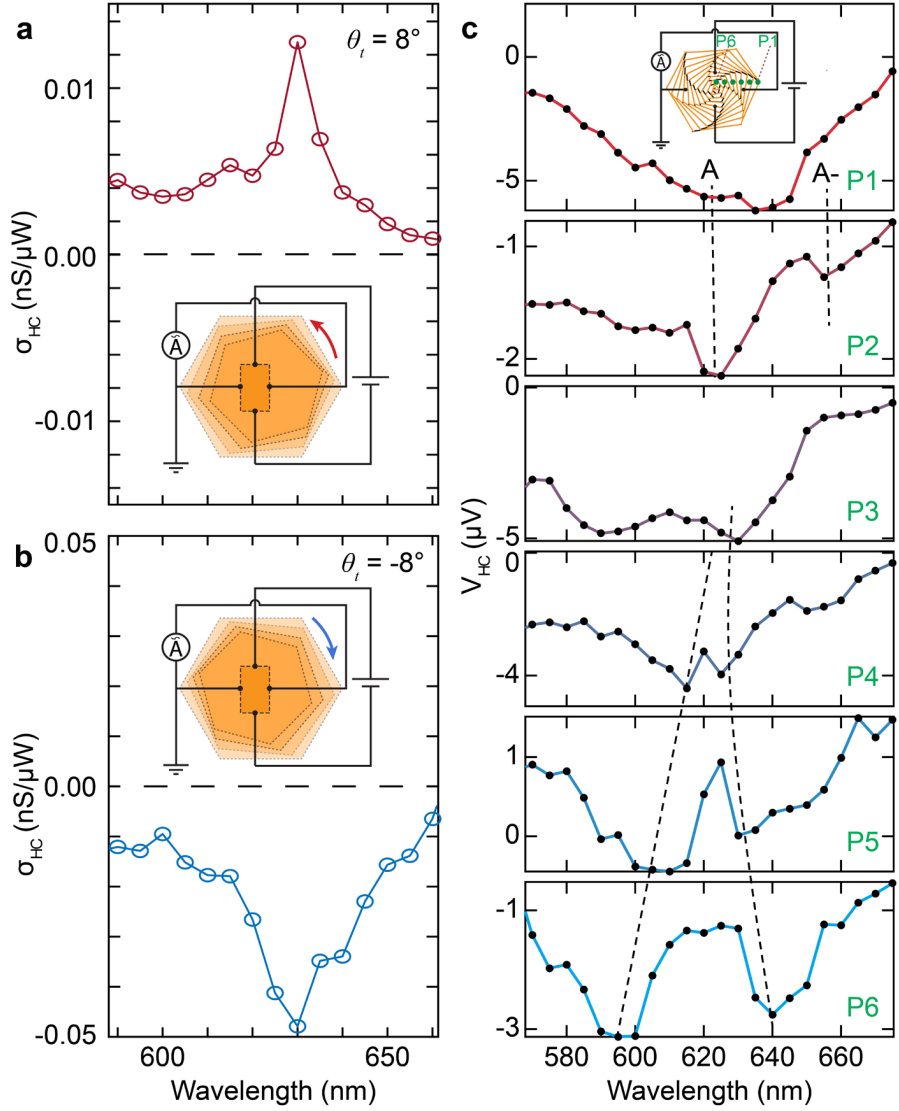

**Fig. 3 Interlinkage between the supertwisted spiral geometry and the opto-twistronic signal.**
**a.** The spectrum of Hall conductivity $\sigma_{HC}$ (C denotes for the part dependent on circular polarization) measured on a ~ $8^o$ left-handed supertwisted sample (D2). Inset: A schematic of device D2. **b.** The spectrum of Hall conductivity $\sigma_{HC}$ measured on a ~ $-8^o$ right-handed supertwisted sample (D3), showing a sign reversal of the response. Inset: A schematic of device D2. **c.** Inset: A schematic of sample D4. Top to bottom: The $V_{HC}$ - wavelength spectra measured at six spots P1- P6 marked on the inset schematics, from the thinnest to the thickest region showing an exciton-photon coupling induced splitting in the Hall voltage.

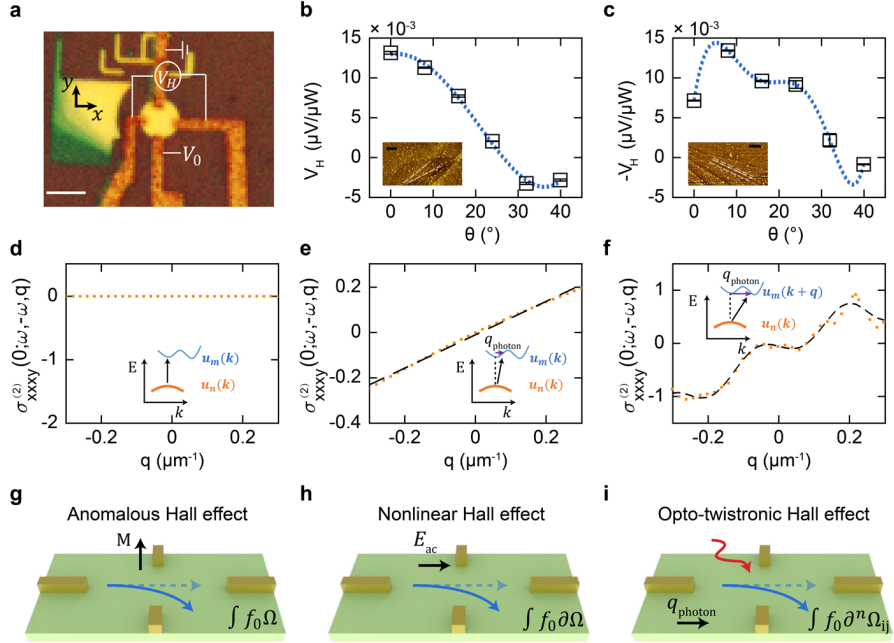

**Fig. 4 Opto-twistronic Hall response from spiral lattice-photon momentum interactions.**
**a.** Optical image of the circular Hall bar device D5. Scale bar: 5 μm. **b.** Incidence angle ($\theta_y$) dependence of $V_{HC}$ on sample D6. Data is presented as black squares with error bar, and the blue dotted line is a polynomial fitting. Scale bar: 1um. Inset: AFM image of sample D6, showing $15-20^o$ twist angles. **c.** Incidence angle ($\theta_y$) dependence of $V_{HC}$ on sample D5. Data is presented as black squares with error bar, and the blue dotted line is a polynomial fitting. Scale bar: 1um. Inset: The AFM image of D5, showing $\sim 1-5^o$ twist angles. **d-f.** A representative second order optical conductivity tensor element calculated for, **d.** monolayer $WS_2$ tight binding model; **e.** prototypical twisted homobilayer TMD model with twist angle $\theta = 1^o$; **f.** supertwisted TMD model with a continuous twist angle $\theta = 2^o$. Insets in **d-f**: Schematics illustrating the corresponding optical excitation processes. Yellow dots are calculated $\sigma^{(2)}_{xxxy}$ and black dashed lines are the polynomial fittings. **g-i.** Schematics illustrating the anomalous Hall effect, nonlinear Hall effect, and photon momentum driven nonlinear Hall effect.